# Ethical AI for Young Digital Citizens: A Call to Action on Privacy Governance

**Author List**
Austin Shouli, Department of Computer Science, Vancouver Island University, Nanaimo, BC, Canada
Ankur Barthwal, Department of Computer Science, Vancouver Island University, Nanaimo, BC, Canada
Molly Campbell, Department of Computer Science, Vancouver Island University, Nanaimo, BC, Canada
Ajay Kumar Shrestha, Department of Computer Science, Vancouver Island University, Nanaimo, BC, Canada

Corresponding author: Ajay K. Shrestha (e-mail: ajay.shrestha@viu.ca).

"This work was supported through the Office of the Privacy Commissioner of Canada's (OPC) Contributions Program award granted to Dr. Ajay Shrestha."

**ACKNOWLEDGEMENT** This project has been funded by the Office of the Privacy Commissioner of Canada (OPC); the views expressed herein are those of the authors and do not necessarily reflect those of the OPC.

**ABSTRACT** The rapid expansion of Artificial Intelligence (AI) in digital platforms used by youth has created significant challenges related to privacy, autonomy, and data protection. While AI-driven personalization offers enhanced user experiences, it often operates without clear ethical boundaries, leaving young digital users vulnerable to data exploitation and algorithmic biases. This paper presents a call to action for ethical AI governance, advocating for a structured framework that ensures youth-centred privacy protections, transparent data practices, and regulatory oversight. We outline key areas requiring urgent intervention, including algorithmic transparency, privacy education, parental data-sharing ethics, and accountability measures. Through this approach, we seek to empower youth with greater control over their digital identities and propose actionable strategies for policymakers, AI developers, and educators to build a fairer and more accountable AI ecosystem.

## I. INTRODUCTION

Artificial Intelligence (AI) has become a crucial component of contemporary digital environments, significantly influencing the interactions of young digital citizens with technology. As AI-driven platforms increasingly facilitate social interactions, educational resources, and digital services, they also introduce complex challenges related to privacy, ethics, and security [1]. These applications often rely on decision-making, significant data collecting, and opaque algorithmic procedures, making it difficult for users, particularly young demographics, to understand how their data is utilized, stored, or monetized [2]. This lack of transparency, accountability, and user autonomy in AI systems has raised significant apprehensions about data privacy, cybersecurity threats, and ethical governance, requiring immediate governmental interventions to protect young users from exploitative AI practices [3].

Despite ongoing global initiatives to regulate AI and implement privacy protection legislation such as the proposed European Union (EU) Artificial Intelligence Act, the EU's General Data Protection Regulation (GDPR), and Canada's Personal Information Protection and Electronic Documents Act (PIPEDA), existing policies frequently neglect the specific needs of young digital users. Studies indicate that many young people lack sufficient knowledge regarding data rights, AI ethics, and cybersecurity protocols, leaving them vulnerable to algorithmic prejudice, targeted monitoring, and possible misuse of their personal information [4]. Moreover, key stakeholders such as parents, educators, and AI professionals, who play a critical role in influencing youth engagement with AI, frequently struggle to navigate the intricacies of AI privacy governance. This has resulted in gaps in AI literacy initiatives, digital safety protocols, and regulatory actions [5]. Given the growing dependence on AI-driven ecosystems, it is essential to build stakeholder-centric policies to ensure the efficient integration of privacy, security, and ethical issues into AI applications affecting young users.

Education is essential in equipping young individuals with the knowledge necessary to make informed decisions about their privacy. However, current AI literacy programs are disjointed, inconsistent, and often inaccessible, lacking structured guidance on data protection, AI transparency, and cybersecurity awareness [6]. AI-driven platforms predominantly prioritize user engagement over fostering privacy-conscious behaviors, increasing the risk of uninformed consent and data exploitation. To address these issues, it is essential to build AI privacy education programs in collaboration with educators, politicians, and AI professionals. Such programs would equip young digital citizens with the necessary skills to navigate AI-powered environments responsibly [7].

The cybersecurity implications of AI, including privacy and education, must be carefully considered. AI technologies introduce novel avenues for cyber threats, such as AI-powered phishing schemes, deepfakes, and vulnerabilities in adversarial machine learning [8]. While traditional cybersecurity measures mostly target corporate and governmental applications, there is a notable gap in cybersecurity strategies tailored to protect young users from risks associated with AI-driven personalization, predictive analytics, and automated decision-making [2]. Enhancing cybersecurity measures in AI applications requires multi-tiered security frameworks, privacy-preserving AI models, and ethical design principles that prioritize the safety of young users without compromising accessibility.

This paper is motivated by the growing disconnect between the widespread integration of AI into youth-facing technologies and the lack of inclusive governance mechanisms that account for the unique privacy risks confronting young digital citizens. While youth are among the most engaged users of AI-enabled systems, they are frequently excluded from the design, regulation, and oversight of these technologies. This research responds to that gap by proposing a stakeholder-driven framework for ethical AI governance that centers youth perspectives on the development of privacy education, regulatory protections, and system design.

In doing so, this paper makes three key contributions. First, it offers a multi-stakeholder analysis of privacy expectations and governance challenges based on the perspectives of young digital citizens, parents, educators, and AI professionals. Second, it identifies specific regulatory and ethical shortcomings in current AI governance, such as insufficient consent protocols, algorithmic opacity, and unchecked data commodification. Third, it presents actionable guidelines and policy recommendations to support youth-centred privacy governance, including proposals for privacy-by-design, digital literacy initiatives, and cross-sector collaboration. This work is distinct in centering youth as active participants in AI governance discourse and in synthesizing stakeholder-specific perspectives that expose structural gaps in current privacy frameworks.

The rest of this paper is organized as follows. Section II reviews the background literature and the evolution of AI privacy concerns. Section III presents stakeholder perspectives and systemic governance challenges. Section IV outlines the best practices and policy guidelines for responsible AI use. Section V concludes with a synthesis of insights and recommendations for building a more ethical, inclusive, and accountable AI ecosystem for young digital citizens.

## II. BACKGROUND AND RELATED WORKS

### *A. The Development of AI Privacy Issues in Digital Environments*

AI has become deeply embedded in digital life, transforming user interactions in social media, healthcare, education, and e-commerce [5]. Although these breakthroughs improve personalization and efficiency, they also present significant privacy problems, especially for young digital citizens who may lack the knowledge or technical skills to navigate intricate AI ecosystems [3]. In contrast to conventional data-driven systems, AI-driven platforms utilize self-learning algorithms that continuously refine data processing methods, affecting content distribution, behavioral forecasting, and automated decision-making [6].

A major difficulty in AI privacy arises from algorithmic opaqueness, commonly referred to as the “black box” effect, which prevents users from understanding the decision-making processes that affect them [1]. This lack of transparency undermines user trust and limits informed consent, as individuals often unknowingly surrender their data without comprehending the ramifications [2]. Generative AI, for instance, collects and analyses vast amounts of personal data, frequently sourced from the internet without explicit consent, raising ethical and legal questions about data ownership and consent validity [4].

Another issue is the growing monetization of personal data, especially inside youth-oriented digital platforms like educational AI tools and social media applications [8]. Research indicates that many AI-based platforms depend on behavioral monitoring and predictive analytics to enhance user engagement and advertising revenue, often with little regulatory oversight [9]. This has resulted in increased concerns over data security, exploitation, and algorithmic bias, wherein AI models perpetuate inequalities based on race, gender, and socioeconomic status [4], [7].

Furthermore, studies utilizing the Privacy Calculus Model (PCM) and Communication Privacy Management (CPM) Theory indicate that users, particularly young digital citizens, weigh perceived benefits like personalization and efficiency against privacy risks such as data misuse or loss of control [10], [11], [12], [13]. However, studies indicate that young people frequently underestimate long-term privacy risks, participating in high-disclosure behaviors without adequately considering the permanence or potential misuse of their data [14]. This demonstrates that young users tend to embrace AI-driven personalization without fully understanding the associated ethical and privacy implications.

Despite these issues, existing legislative frameworks remain predominantly reactive, struggling to adapt to the rapid advancements in AI [8]. Many policies focus on adult privacy issues, neglecting the unique socio-cognitive factors that influence youth privacy practices and risk assessments [15]. Moreover, studies about AI privacy governance often prioritize legal compliance over user-centric transparency, limiting young users' capacity to effectively manage their privacy settings [5].

This study aims to address current gaps by examining how young digital citizens, parents, educators, and AI professionals perceive privacy in AI-driven contexts from a multi-stakeholder perspective. Unlike previous research, which often examines privacy through a generalized regulatory lens, this study focuses on the unique challenges faced by young users, emphasizing the need for proactive, inclusive, and ethically responsible AI policies that align with user expectations and evolving technological advancements.

### *B. Stakeholder Perspectives on Privacy in AI Systems*

AI governance is influenced by the concerns, expectations, and interactions of various stakeholders, including young digital citizens, parents, educators, and AI professionals. Each group fulfils a unique function in shaping privacy standards, regulatory focuses, and ethical AI frameworks. However, current research indicates that variations in privacy expectations among different groups frequently result in fragmented governance strategies and errors in policy execution [2]. Young digital citizens, for example, frequently interact with AI-driven platforms, yet they often possess inadequate knowledge of how their personal data is processed, resulting in a paradox where they voice privacy concerns yet engage in high-risk online behaviors [16]. Research shows that young people often trade privacy for convenience, failing to recognize the long-term risks of data sharing, behavioral monitoring, and algorithmic profiling [1]. AI-driven systems, such as recommendation engines, continuously gather user data, often without explicit consent protocols, prompting concerns over data autonomy and transparency [4]. Similarly, educational platforms, digital assistants, and social media networks accumulate extensive personal data while providing minimal openness concerning data retention, third-party access, and monetization [14].

Parents and educators play a crucial role in influencing the privacy behaviors of young users, however, their viewpoints frequently conflict with those of the younger generation. Numerous parents support enhanced privacy regulations, concerned that unregulated AI interactions could subject adolescents to privacy violations, misinformation, and manipulation [17]. Nonetheless, excessive parental monitoring can sometimes undermine trust and autonomy [9]. At the institutional level, educational institutions are beginning to incorporate AI literacy into the curricula but may lack structured programs that sufficiently address the privacy risks linked to AI [15]. Collaborative approaches, which involve parents and educators engaging youth in dialogue instead of enforcing unilateral restrictions, have demonstrated potential; however, current educational frameworks remain underdeveloped, resulting in a significant deficiency in AI ethics instruction [10].

AI professionals, including developers, engineers, and policymakers, approach privacy from a technical and governance perspective, focusing on efficiency, security, and compliance rather than user-centric issues. While many AI engineers recognize the ethical ramifications of automated decision-making, especially concerning algorithmic bias, transparency, and data protection, there remains a gap between technological advancement and the ethical expectations of end users, especially young digital citizens [1], [8]. A significant obstacle in AI governance is the lack of standardized privacy benchmarks, as frameworks like the GDPR and the Office of the Privacy Commissioner of Canada (OPC) guidelines provide foundational principles but are inconsistently enforced across platforms [4]. Research highlights the excessive dependence on self-regulation by technology companies, where profit-driven algorithmic enhancements overshadow ethical protections [2]. Additionally, discussions on AI explainability persist without resolution, with engineers prioritizing technical feasibility while end-users seek more transparent explanations for algorithmic decisions [7]. This disconnect highlights the need for enhanced cross-sector collaboration, guaranteeing that AI governance encompasses not only corporate interests but also the ethical standards of all stakeholders.

The differences in privacy expectations across stakeholders reveal a fundamental misalignment: youth prioritize autonomy, parents emphasize protection, and AI professionals focus on efficiency and compliance. This fragmentation undermines privacy governance initiatives, leading to policy inconsistencies, restricted user control, and regulatory gaps [5]. A multi-stakeholder governance framework is essential to align privacy safeguards with the realities of AI ecosystems. Research indicates that participatory policymaking, involving young users, parents, educators, and AI developers in the formulation of privacy guidelines, promotes more flexible and ethical governance [4]. Transparent data-sharing rules, improved AI literacy initiatives, and enforceable privacy regulations are crucial for establishing a more equitable AI environment that emphasizes user rights, ethical innovation, and sustained accountability [6]. This research addresses stakeholder-specific challenges through structured policy discussions, contributing to the development of ethical, inclusive, and enforceable AI privacy norms that balance autonomy, transparency, and regulatory control.

### *C. Regulatory and Ethical Challenges in AI Privacy Governance*

Privacy governance in AI-driven environments poses significant ethical and policy challenges owing to the evolving characteristics of AI technology and the growing dependence on algorithmic decision-making. Although current data protection frameworks, including the GDPR and directives from the OPC [4], aim to protect personal data, they often fall short in addressing the emerging privacy risks associated with AI systems [1], [18]. A significant limitation of these frameworks is their reactive nature, as legal standards struggle to keep pace with advancements in AI-driven data processing and predictive analytics [5]. Effective AI privacy governance must evolve to address issues like transparency, user control, and algorithmic accountability [4], [7].

*1) Consent and Autonomy*

A fundamental ethical concern in AI governance is the lack of meaningful user consent, especially among young digital citizens who engage with AI-driven platforms without fully comprehending the extent of data collection and processing [19]. Numerous AI systems function under ambiguous consent protocols, where users unknowingly agree to massive data collection through lengthy and intricate terms of service agreements [20], [21]. Research indicates that young users in particular struggle to critically assess AI privacy policies and engage with digital platforms without considering the long-term consequences of data sharing [22], [23]. The Privacy Calculus Model (PCM) suggests that individuals evaluate the risks and benefits of data disclosure prior to making privacy decisions [10], [13], yet studies reveal that young users often overlook long-term privacy risks in favour of immediate conveniences [14], [24].

*2) Algorithmic Profiling and Accountability*

Another significant regulatory concern is ensuring accountability in AI systems, especially when platforms use predictive profiling, behavioral tracking, and targeted advertising without users' explicit awareness [25], [26]. The OPC expresses apprehensions over AI systems using vast datasets sourced from the internet [4], which frequently lack transparent permission protocols, exacerbating the risks of data commodification and surveillance [18]. Automated decision-making, which powers content recommendations, personalized services, and even financial decisions, often operates without transparency, limiting user autonomy and understanding of AI-generated reasoning [22], [27], [28], [29].

*3) Bias and Discrimination*

Algorithmic prejudice and discrimination represent significant risks to privacy governance, as AI models often perpetuate societal inequalities present in their training datasets [30], [31]. AI systems used in hiring, education, and social media exhibit patterns of gender, racial, and socioeconomic prejudice, resulting in discriminatory consequences [32], [33]. Proposed bias mitigation solutions, including fairness-aware machine learning and ethical AI audits, are inconsistently implemented across digital platforms [34], [35]. Regulatory bodies face challenges in creating uniform accountability frameworks that address fairness, transparency, and user rights in algorithmic design [26], [36].

*4) Data Sharing and Surveillance Risks*

Third-party data sharing and corporate surveillance also pose ethical challenges, as AI-driven companies often treat personal data as a commodity for advertising, behavioral analysis, and financial gain [4], [37]. Research indicates that platforms in social media, education, and healthcare exploit user data for profit through undisclosed arrangements with advertisers and data brokers, often evading regulatory oversight [23], [28]. This lack of transparency in data transactions erodes user trust, particularly among young digital citizens, who are often unaware of how their personal information is used beyond the principal platform [9], [38]. Research suggests that user confidence in AI systems diminishes when transparency is lacking, highlighting the necessity for enhanced data protection regulations and algorithmic explainability [18], [39].

*5) Education and Literacy Gaps*

Initiatives to improve AI privacy governance include broadening AI literacy programs to educate users on data protection rights, consent protocols, and algorithmic accountability [19], [40]. Interactive privacy education, youth-focused AI literacy programs, and parental digital literacy training can substantially improve user knowledge of privacy risks [29], [41]. However, current educational approaches often neglect to address practical AI privacy scenarios, leaving users inadequately equipped to manage intricate digital environments [30]. To address these issues, AI privacy regulations must incorporate the perspectives of young users, ensuring that policies align with their expectations and dynamic digital behaviors [18], [42].

*6) Toward Participatory and Enforceable Governance*

Stakeholder-driven governance frameworks have been suggested to address these challenges, promoting multi-stakeholder involvement in AI policy development and incorporating perspectives from young users, parents, educators, and AI experts [18], [43]. Research highlights the need for youth involvement in AI legislation, as ongoing discussions often exclude young digital citizens, despite their considerable engagement with AI technologies [20], [26]. Furthermore, enforceable AI privacy policies, such as standardized opt-out mechanisms, enhanced consent protocols, and regulatory scrutiny of algorithmic decision-making, are crucial for creating ethical AI frameworks that emphasize user autonomy, digital rights, and privacy protection [4], [35].

In conclusion, AI privacy governance must transition from reactive compliance to proactive, user-centric approaches that tackle the unique privacy issues presented by AI systems [9], [18]. Key policy recommendations include a stakeholder-driven, enforceable governance framework that aligns privacy policies with the digital realities of young users, ensuring ethical AI development that emphasizes autonomy, transparency, and sustained data protection [26], [42]. This study contributes to policy discussions by advocating for inclusive, enforceable, and adaptive privacy policies that safeguard young digital citizens in AI-driven environments.

### *D. Gaps in Existing AI Privacy Policies and the Need for Stakeholder-Inclusive Governance*

Current AI privacy regulations are disjointed and predominantly reactive, failing to keep up with rapid technological advancements and the changing dynamics of data privacy issues. While regulatory frameworks like the GDPR in Europe and the California Consumer Privacy Act (CCPA) in the United States offer essential guidelines, they often fail to address the specific vulnerabilities of young digital citizens who interact with AI-driven platforms daily [4], [18]. The OPC has highlighted shortcomings in existing privacy legislation, citing ineffective consent mechanisms, opaque algorithmic processes, and the absence of legal safeguards for youth engaging with AI ecosystems [23]. Despite growing recognition of these constraints, regulatory efforts continue to prioritize corporate compliance over user-centric governance, leading to privacy policies that limit young users' control over their personal data [38], [42].

A major gap in current regulations is the absence of explicit and enforceable standards regarding AI-driven data profiling, behavioral tracking, and automated decision-making [28], [40]. AI systems increasingly depend on machine learning models to infer user preferences, emotions, and behaviors, yet there are few safeguards guaranteeing transparency or user control over these deductions [9], [24]. While the GDPR enforces transparency in data processing, it does not explicitly govern algorithmic profiling, which influences digital experiences through targeted advertising, content curation, and predictive analytics [1], [7]. Consequently, young digital citizens utilize AI-driven applications without comprehending the impact of algorithmic decision-making on their online interactions or the aggregation and monetization of their behaviors [26], [44]. The lack of strong safeguards against profiling raises issues around privacy degradation, data commercialization, and the enduring effects of tailored digital environments on young people's autonomy.

Another significant gap in AI privacy regulations is the reliance on inadequate consent mechanisms. Informed consent is fundamental to data protection legislation; however, current frameworks fail to address the cognitive and behavioral disparities that affect how young users interact with AI technologies [20]. Studies show that the majority of young users rarely examine privacy agreements thoroughly and frequently consent to conditions without comprehending the ramifications of their decisions [22], [23] This has resulted in what experts term the "consent paradox," where users appear to give consent but lack a clear understating of how their data is gathered, used, or shared [18]. AI-driven platforms exacerbate this problem by integrating "dark patterns" into their design-manipulative interface strategies that discreetly encourage users to accept privacy-invasive settings [35], [43]. Despite extensive advocacy for the creation of more accessible, interactive, and age-appropriate consent frameworks, privacy policies persist in prioritizing legal formalities over enabling users to make informed decisions regarding their data [29], [30].

Cross-platform data sharing presents another regulatory challenge, as AI-driven applications in social media, education, and healthcare, often engage in extensive data-sharing practices that are poorly understood by end users [34], [38]. Personal data gathered on one platform is frequently combined with datasets from other platforms, allowing AI models to create comprehensive behavioral profiles without user awareness or consent [4], [18]. For example, AI-driven educational tools amass extensive student data, encompassing learning behaviors, cognitive patterns, and emotional responses, while offering minimal transparency regarding the storage, sharing, or application of this data beyond the classroom [27], [32]. Similar concerns arise in AI-driven healthcare applications, as patient data is frequently utilized for algorithmic predictions without adequate protection against misuse, data breaches, or unexpected repurposing [28], [41]. The lack of enforceable regulations on cross-platform data sharing poses considerable risks for young users, who are often unaware of how widely their personal information is circulated.

A further issue is the lack of youth representation in AI governance. Privacy regulations often adopt an adult-centric viewpoint, neglecting the social, cognitive, and digital literacy factors that influence young users' interactions with AI [20], [45]. Research shows that youth-specific privacy concerns, such as peer influence, changing risk perceptions, and a desire for digital autonomy, are rarely considered in policy discussions [26], [42]. The prevailing emphasis on legal compliance over user education and participatory governance leaves young users with little say in the privacy standards that directly affect them [40], [46]**.** The lack of representation leads to policies that inadequately address the real privacy concerns faced by young digital citizens, limiting their ability to advocate for their privacy rights in AI-driven environments.

This paper addresses these gaps by advancing a stakeholder-inclusive perspective that foregrounds youth not simply as vulnerable users but as active agents whose lived experiences reveal systemic limitations in current AI privacy governance. Unlike prior studies that frame youth privacy in isolation or subsume it within adult-focused frameworks, our work analyzes how young users, parents, educators, and AI professionals differentially understand and respond to privacy risks. Scholars such as Livingstone and Third have emphasized the need to shift from protectionist discourses to enabling rights-based approaches that respect youth agency in digital contexts [47]. Building on this, our study integrates perspectives from young digital citizens, parents, educators, and AI professionals to provide a more granular and contextualized account of privacy expectations, regulatory misalignments, and lived experiences with AI systems.

Whereas much of the AI governance literature isolates privacy risks as either legal or individual [5], [8], [9], our framework connects youth privacy to broader systemic failures in policy, education, and technological design. Previous studies have underscored the limitations of abstract ethical principles and the risks of industry-led self-regulation [4], [18], [48], [49], yet few offer a practical model that bridges stakeholder expectations, developmental needs, and governance strategy. Our contribution is distinct in combining empirical insight with normative direction, advancing a youth-centred governance framework that aligns with the United Nations Convention on the Rights of the Child's digital rights agenda [50].

In doing so, this research moves beyond surface-level consultations or age-specific safeguards to propose a participatory model of AI governance that is responsive to the structural asymmetries young users face in algorithmic environments. By synthesizing stakeholder-specific insights, we illustrate how privacy expectations and governance priorities vary significantly among youth, parents, educators, and AI professionals, and how those tensions may be reconciled through cross-sector collaboration and shared accountability [6], [10], [41]. This approach responds directly to the policy fragmentation and institutional inertia that often sideline youth perspectives in favor of legal compliance or adult gatekeeping [20], [26], [42], [51].

Our findings also demonstrate the need for a governance paradigm that connects ethical design, education, and enforceable safeguards, rather than treating them as discrete policy arenas [4], [9], [18], [41]. By situating youth privacy within the broader context of institutional responsibility, consent culture, and participatory design, this study contributes a distinct framework that advances beyond descriptive analysis and moves toward practical, youth-centered AI policy development. The results highlight the immediate necessity for enhanced regulatory measures, including simplified consent processes, stricter controls on behavioral profiling, greater algorithmic transparency, and comprehensive AI literacy programs tailored to young users [20], [24]. This research promotes participatory AI governance models that involve young digital citizens in policy conversations, ensuring their perspectives, experiences, and expectations are integrated into privacy frameworks [29], [46]. Enforceable regulations are essential to hold AI platforms accountable for ethical data use, limit surveillance, and advance user-centric privacy protections[40], [43]. This paper proposes recommendations for a youth-centred AI privacy governance model that harmonizes regulatory frameworks with the digital realities of young users, ensuring that AI technologies prioritize privacy rights, ethical innovation, and regulatory accountability in a dynamic digital landscape [4], [18].

To synthesize the limitations identified across current literature, Table I summarizes the key research gaps in existing AI privacy governance frameworks. These gaps span regulatory, ethical, educational, and participatory dimensions, highlighting the disconnect between rapidly evolving AI technologies and the insufficient protections afforded to young digital citizens. The table contextualizes each gap with representative references, reinforcing the need for a youth-centred, stakeholder-inclusive approach that moves beyond compliance toward proactive and ethical AI governance.

TABLE I
SUMMARY OF KEY RESEARCH GAPS IN AI PRIVACY GOVERNANCE LITERATURE

| Research Gap | Description | References |
|---|---|---|
| Lack of Youth-Specific Policy Frameworks | Existing regulations primarily address adult privacy concerns and don't incorporate comprehensive youth-specific cognitive, social, and digital literacy factors. | [4], [18], [20], [23], [26], [38], [42], [45] |
| Inadequate Consent Mechanisms | Consent processes are often complex and inaccessible, particularly for young users, leading to the "consent paradox" and manipulation through dark patterns. | [18], [20], [22], [23], [28], [30], [35], [43] |
| Insufficient Regulation of Algorithmic Profiling | Current frameworks lack enforceable standards on behavioral profiling, automated decision-making, and targeted content curation. | [1], [7], [9], [24], [26], [27], [40], [44] |
| Opaque Data Sharing Practices | AI systems engage in cross-platform data sharing without sufficient transparency or user control, especially in critical sectors including education. | [4], [18], [27], [29], [32], [34], [38], [41] |
| Limited Youth Participation in Governance | Youth perspectives are often excluded from policymaking, resulting in protectionist approaches that limit digital agencies and misalign with their lived experiences. | [10], [20], [26], [40], [42], [46], [47], [51] |
| Overreliance on Industry Self-Regulation | Ethical design is often subordinated to corporate interests, with companies setting their own standards for AI transparency and accountability. | [2], [4], [8], [18], [35], [48], [49] |
| Fragmentation Across Stakeholder Priorities | Diverging expectations between youth (autonomy), parents (protection), educators (awareness), and professionals (compliance) hinder unified governance. | [1], [4], [5], [6], [8], [9], [14], [16], [17] |
| AI Literacy and Education Gaps | Existing educational initiatives lack structured, context-specific instruction on AI privacy risks, particularly for young digital citizens. | [10], [19], [28], [30], [40], [41], [42] |

## III. DISCUSSION

### *A. Stakeholder-centric Privacy regulation in AI applications*

As artificial intelligence technology becomes increasingly prevalent in daily life, there has been increased discussion around the need for regulation and regulatory approaches. While the European Union has traditionally led the way in regulating technology, Canada and the United States tend to lag behind. In 2018, the GDPR went into effect across the EU, providing individual data protection rights and requirements for data processing and handling corporations. While Canada does have the Personal Information Protection and Electronic Documents Act (PIPEDA), this policy covers a broader range of data rights, and the recent progress in AI technology has necessitated regulation and policy that accounts for the new capabilities of these tools.

Prior research on online youth privacy has indicated that young people, parents, and educators largely support increased regulation and oversight of AI technology as it pertains to young digital citizens [13]. Specifically, young digital citizens expressed a need for privacy tools they can understand, highlighting a need for platforms and services using AI to implement clear and interpretable privacy controls. Young users are often asked to agree to dense and lengthy user agreements, with little opportunity for privacy or data collection control.

Furthermore, many young digital citizens felt that their informed consent was not sought out, or was circumvented, leaving users unsure of what data was being collected, and how it was being stored, used and sold. Platforms that specifically engage with and encourage young users should be required to implement interpretable privacy policies and accessible privacy controls, allowing young users and their guardians to manage their data privacy in an informed and consensual manner.

*B. Educational Initiatives*

In today's fast-paced world, technological advancements have the potential to outstrip the ability of users to fully understand the implications of using this technology. In a world of lengthy user agreements and opaque corporate policies, it is becoming ever more difficult to manage and track one's online footprint and understand what data is collected and why. Recent studies show that a primary concern among young digital citizens regarding data privacy and AI technology is a lack of understanding regarding what data is being collected, and what it is being used for. In a recent survey, multiple respondents expressed the sentiment that they "do not know enough about them to be concerned about what I should be concerned about" [13]. With the rapid pace of AI development in the last few years, it is no surprise that users are finding they do not have the prerequisite knowledge to understand what data privacy they should be concerned about. In the same survey, older respondents, including parents and educators, expressed similar sentiments. Throughout this research, it has become clear that education and awareness of privacy issues have failed to keep pace with technological development, and young digital citizens are engaging with online resources without an adequate understanding of the long-term risks involved.

In the same study, while some educators made an effort to incorporate AI tools in the classroom, this was a minority group, and there was no formal system for engaging with AI tools. The educators who did incorporate such technology did so of their own volition, and with no established framework for education. While such teachers should be lauded for attempting to close the technological gap, a formal education framework for approaching AI should be explored at the federal, or at least provincial level, in Canada. Many Canadian students are already utilizing AI tools in their homework and assignments, in ways that may violate academic integrity policies [13]. Other online tools, such as social media, are gathering data on young users without the user's understanding of how that data may be stored and used. Data Privacy and AI education should be incorporated into the curriculum in Canadian schools, providing young digital citizens with an introduction to the surveillance capitalistic system that dominates our economy, and providing youth with a basis for making informed data decisions in their online behavior.

To make these efforts practical, we recommend curriculum models that emphasize active learning, co-creation, and interdisciplinary integration. For example, project-based learning modules can engage students in real-world scenarios involving algorithmic bias, targeted advertising, or data-sharing practices. Lessons can be scaffolded by age group and designed in collaboration with students, educators, and privacy experts to ensure relevance and accessibility. Pedagogical approaches might include case study analysis, student-led privacy audits of school-used apps, or simulations of AI decision-making processes to illustrate trade-offs between personalization and privacy. Such approaches can also be embedded across subjects, from digital civics and social studies to computer science and health education. Importantly, educator training must be a central component of this effort, equipping teachers with the foundational knowledge and resources to confidently guide classroom discussions on ethical AI and digital rights.

To further clarify the interconnected roles of various actors, Figure 1 presents a visual taxonomy mapping five core stakeholder groups: youth, parents, educators, AI professionals, and policymakers, onto key governance dimensions in AI privacy including autonomy, protection, transparency, digital literacy, and regulatory oversight. This taxonomy reveals how each group contributes to shaping privacy norms, identifying gaps in current governance strategies, and enabling ethically sound digital ecosystems. For instance, while youth are often positioned as data subjects, their evolving agency as digital citizens warrants explicit attention to autonomy and literacy. Educators and parents function as mediators of digital awareness, while AI professionals and policymakers shape the technological and regulatory infrastructure underpinning privacy protections.

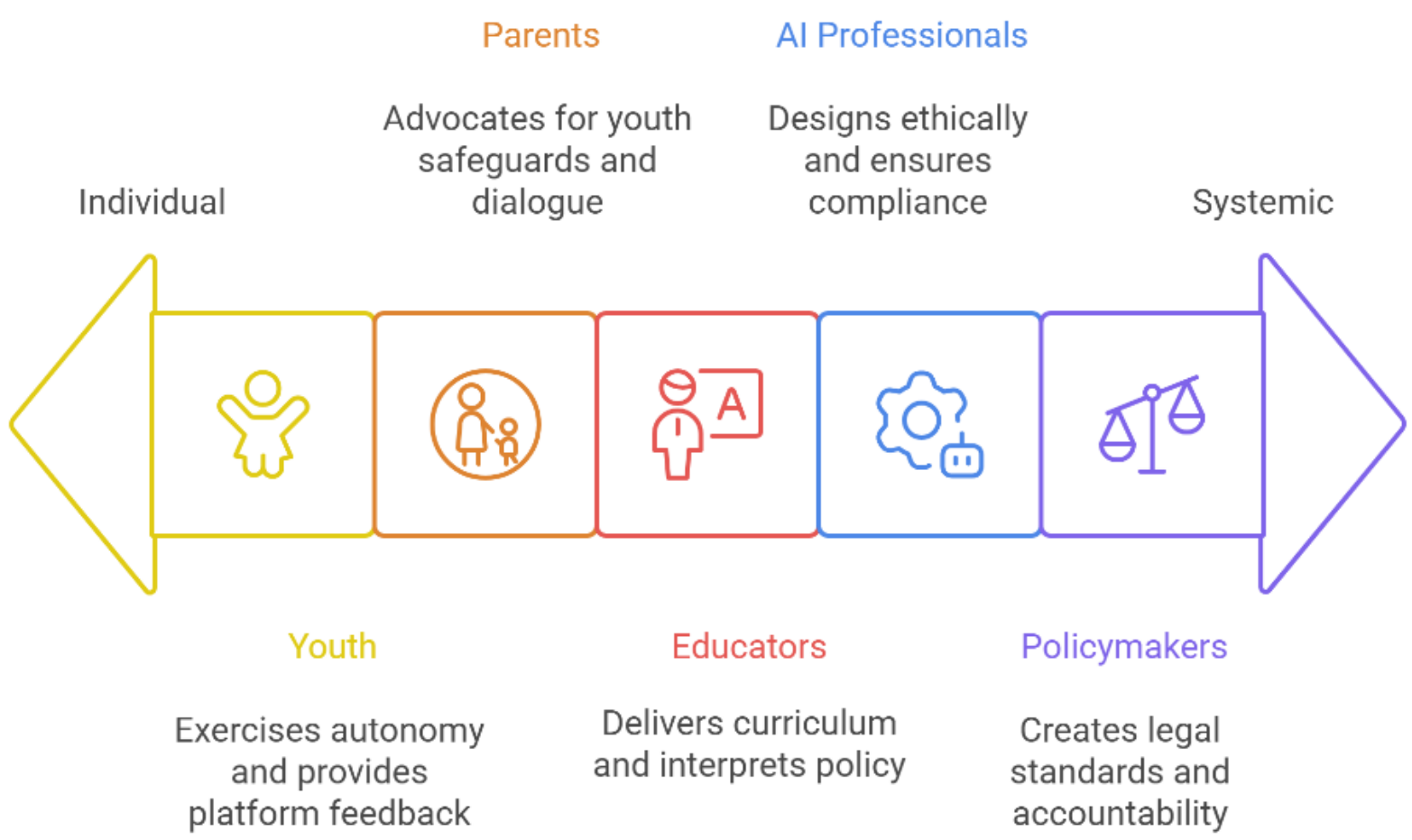

FIGURE I STAKEHOLDER ROLES IN AI PRIVACY GOVERNANCE, FROM INDIVIDUAL TO SYSTEMIC

### C. *Cybersecurity Initiatives*

Strong cybersecurity safeguards are now more important than ever as AI systems continue to be incorporated into vital industries like social media, healthcare, and education. Large volumes of private user data are processed by AI-driven apps, frequently with no user control or transparency. As a result, these apps are vulnerable to cyber threats like algorithmic manipulation, deepfake assaults, and AI-powered phishing. Young digital citizens, who are more prone to interact with AI applications without fully comprehending the security dangers involved, face special difficulties as a result of the changing landscape of AI-driven cyber threats. The protections intended for younger users, whose data is constantly monitored, collected, and analyzed in ways they may not understand, are severely lacking in current cybersecurity frameworks, which frequently concentrate on corporate and governmental protection leaving a significant gap in protections designed for younger users, whose data is continuously tracked, aggregated, and analyzed in ways they may not be aware of.

Data vulnerability brought on by cross-platform data sharing and inadequate encryption mechanisms is one of the main cybersecurity issues in AI ecosystems. Because AI-powered customization mostly depends on behavioral analysis and predictive modelling, security lapses on one platform might have a ripple effect on other AI-driven platforms. Furthermore, there are serious concerns associated with the absence of defined cybersecurity standards for AI decision-making since adversarial attacks can tamper with AI outputs, jeopardize automated recommendations, and spread false information. A multi-tiered security strategy is needed to address these threats, which includes blockchain-based safe data-sharing methods, differential privacy strategies, and federated learning for privacy-preserving AI. In addition to being a technological requirement, enhancing cybersecurity in AI systems is also morally required, since it guarantees that data protection policies are in line with the rights and expectations of young digital citizens.

A youth-centric cybersecurity approach must be incorporated into future policy frameworks, including AI literacy initiatives that teach young people about data encryption awareness, digital self-defence, and safe online conduct. By incorporating security measures into AI models directly rather than as an afterthought, developers must give privacy-by-design principles priority when creating AI. Legal frameworks must also change to hold AI platforms responsible for cybersecurity lapses and to clearly define roles for firms that handle data on children. To mitigate AI-driven cyber threats and build a more secure and resilient AI ecosystem for future generations, it will be crucial to adopt worldwide AI security standards and enforce data protection laws more strictly.

### D. *Risks to Sensitive Areas of Research in AI-Driven Systems*

The integration of AI technologies into research infrastructures presents emerging risks, particularly in sensitive domains such as healthcare, education, behavioral science, genomics, and mental health. These fields often involve highly personal and context-specific data, making them particularly vulnerable to misuse, unauthorized disclosure, or unintended profiling. Modern data-intensive systems utilize AI models that can infer sensitive characteristics from minimal inputs, even when data has been de-identified. As recent studies demonstrate, machine learning techniques can re-identify individuals by correlating anonymized datasets with auxiliary information, undermining conventional privacy safeguards [52]. This raises serious concerns about the adequacy of existing ethical review processes, especially when research participants, such as young digital citizens are unable to meaningfully consent to how their data may be processed, shared, or monetized.

Cloud-based research platforms and federated learning architectures, while offering scalability and collaboration, also introduce new governance challenges. Sensitive data is often hosted by third-party providers or processed across international jurisdictions, where legal protections may be inconsistent or difficult to enforce [53]. This fragmentation can lead to compliance gaps in regulations like the GDPR or Canada's PIPEDA, particularly when data flows across sectors or borders. In addition, commercial cloud services used in academic settings may default to terms of service that permit data aggregation or secondary use beyond the scope of research ethics approvals, exposing participant data to profiling, algorithmic discrimination, or commercial exploitation [49].

These risks are compounded in youth-focused studies, where developmental, cognitive, and social vulnerabilities necessitate enhanced safeguards. However, few AI research frameworks integrate youth-specific privacy measures or participatory data governance models that account for their evolving capacities and digital rights [47]. Without enforceable technical and procedural standards for privacy-preserving research, such as differential privacy, homomorphic encryption, or transparency logs, researchers may inadvertently expose participants to long-term harms through the very infrastructures intended to support ethical inquiry. A forward-looking approach must incorporate privacy-by-design principles into research workflows, mandate institutional accountability for AI model explainability, and encourage multi-stakeholder oversight that includes youth voices in setting ethical standards for sensitive research contexts.

To contextualize these risks within a proactive framework for intervention, Figure 2 presents a conceptual model that centers young digital citizens within a secure and layered governance ecosystem. Represented visually as a lock, the model highlights the interplay of seven interdependent levers required to safeguard youth privacy in AI systems.

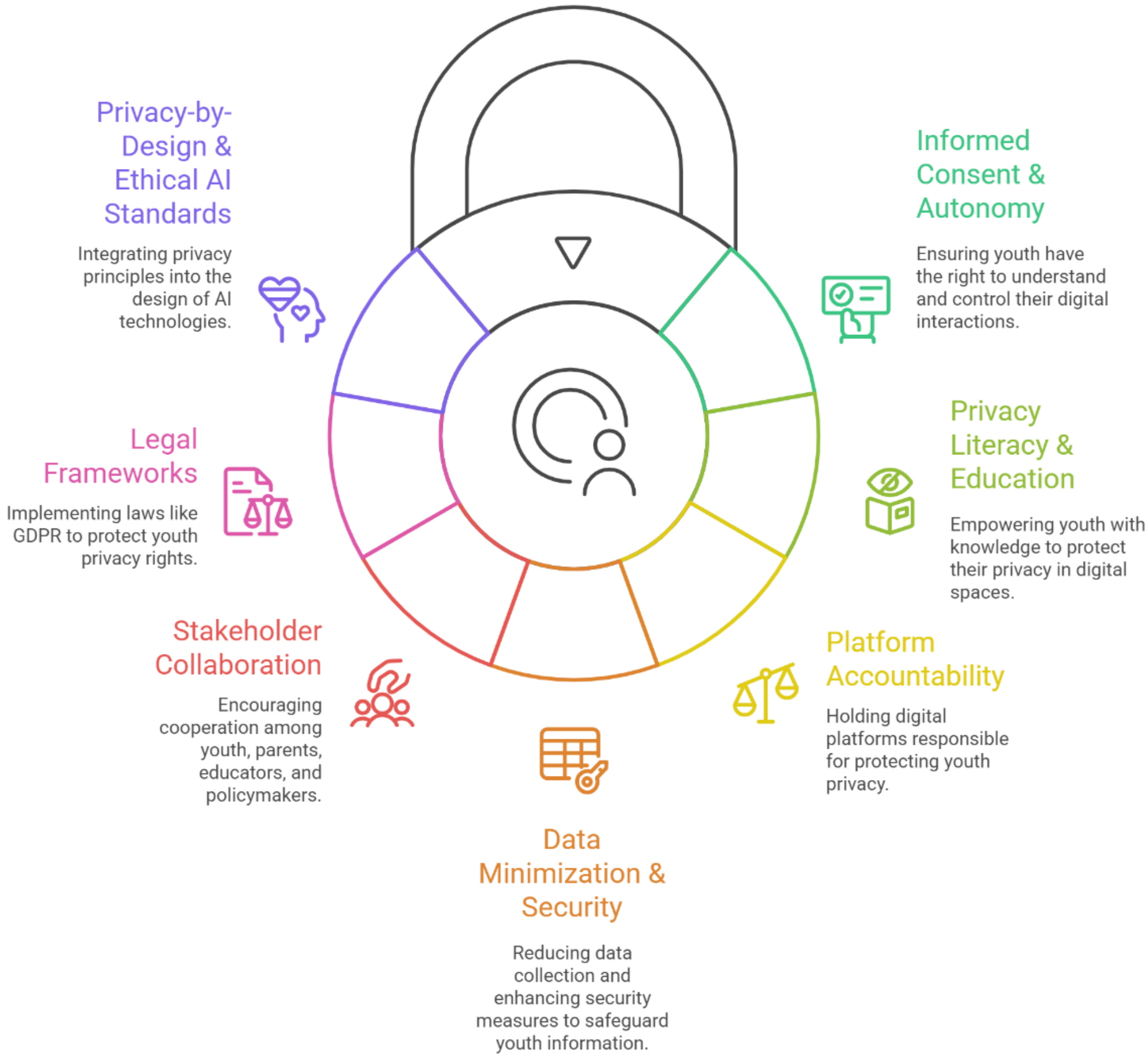

FIGURE 2 CONCEPTUAL FRAMEWORK FOR YOUTH AI PRIVACY

## IV. GUIDELINES AND POLICY RECOMMENDATIONS

To address the identified concerns from the stakeholder groups, a set of draft guidelines was developed, emphasizing practicality, inclusivity, and adaptability across various AI contexts. These guidelines aim to provide actionable strategies for parents, educators, AI professionals, and youth to ensure responsible AI privacy management. The following sections outline the best practices, step-by-step guides, and proposed policy recommendations, while also addressing challenges and future considerations.

### *A. Best Practices*

#### *1) Informed Consent and Age-Appropriate Notices*

One of the foundational elements of ethical AI governance is ensuring that users, particularly young digital citizens, are fully informed about how their data is collected, stored, and used. Clear and straightforward consent forms should be provided, explaining the purpose of data collection and its intended use. To improve comprehension, especially among younger audiences, a mix of text, visuals, and short videos should be employed. Additionally, ongoing awareness initiatives should prompt youth to periodically review their data-sharing preferences and privacy settings, ensuring that they remain in control of their digital footprint.

#### *2) Privacy by Design*

AI systems should be built with privacy at the forefront of their design. This includes minimizing data collection to only what is essential for the application functionality. Ensuring secure storage and transfer of data through encryption and providing user-friendly tools such as privacy dashboards that allow users to easily delete or download their data. Where possible, identifiable data should be replaced with anonymized or pseudonymized versions to enhance privacy without compromising AI functionality. Privacy Impact Assessments (PIAs) should be conducted to evaluate how AI systems handle data, identify potential risks, and implement necessary safeguards.

#### *3) Transparency and Explainability*

Transparency is critical in building trust between users and AI systems. Privacy policies should be written in plain, jargon-free language, making it more accessible for users to understand how their data is being handled. Algorithmic transparency is equally important, with general explanations provided about how AI systems make decisions, particularly in educational or recommendation systems. Regular feedback loops involving students, parents, and educators should be established to gather input on the perceived fairness and safety of AI applications, ensuring that these systems evolve in a way that aligns with user expectations.

#### *4) Monitoring and Accountability*

To maintain the integrity of AI systems, regular internal audits should be conducted to identify and address data leaks, biases, or security vulnerabilities. Users should be encouraged to report privacy concerns or AI errors through dedicated reporting mechanisms, such as helpdesks or online forms. Independent oversight, involving neutral experts, ethicists, or advisory groups, should be engaged to evaluate the impact of AI on youth privacy and fairness. Clear ethical data handling policies must be established to ensure that AI systems comply with privacy regulations and ethical guidelines.

#### *5) Education and Literacy*

Equipping young digital citizens with a foundational knowledge of data privacy and ethical AI is essential for long-term digital empowerment. Schools should integrate AI-related privacy education across disciplines, such as social studies, digital civics, and computer science, to ensure a broad and contextual understanding. In parallel, teachers must receive appropriate training and resources to confidently guide classroom discussions on algorithmic risks, data protection, and surveillance culture. Student-led activities, including privacy audits, digital ethics case studies, and project-based learning modules, can foster critical thinking and responsible online behavior. These efforts must be scaffolded by age and supported by interdisciplinary collaboration.

#### *6) Multi-Stakeholder Collaboration*

The development of AI policies should be a collaborative process among policymakers, educators, AI developers, and youth representatives. This ensures that AI applications are designed with youth privacy in mind, prioritizing fairness, security, and informed consent. Schools and institutions can play a pivotal role in advocating for youth-centric AI policies, fostering an environment where ethical considerations are at the forefront of technological innovation.

To support the implementation of these best practices, Table II presents an actionable checklist for youth-centered AI privacy governance. Organized across six thematic domains, the checklist serves as a practical reference for educators, policymakers, and developers. It is designed to translate conceptual principles into concrete actions that can guide policy development, curriculum planning, and ethical AI system design.

TABLE II
ACTIONABLE CHECKLIST FOR YOUTH-CENTERED AI PRIVACY GOVERNANCE

| Category | Recommended Action Items |
|---|---|
| 1. Informed Consent | ☐ Provide age-appropriate, visually supported consent forms<br>☐ Explain data collection purposes clearly and concisely<br>☐ Include periodic prompts to revisit privacy preferences |
| 2. Privacy by Design | ☐ Limit data collection to what is strictly necessary<br>☐ Use encryption and anonymization to protect personal data<br>☐ Implement privacy dashboards for youth and guardians<br>☐ Conduct regular Privacy Impact Assessments (PIAs) |
| 3. Transparency | ☐ Use plain language in privacy policies and AI descriptions<br>☐ Provide users with general explanations of how AI decisions are made<br>☐ Enable feedback loops with students, parents, and educators |
| 4. Monitoring and Accountability | ☐ Conduct internal audits for bias, privacy violations, and security flaws<br>☐ Establish accessible reporting mechanisms for users<br>☐ Include independent ethical oversight (e.g., advisory boards or expert panels) |
| 5. Education and Literacy | ☐ Integrate AI privacy topics into school curricula across disciplines<br>☐ Provide teacher training on ethical AI and privacy literacy<br>☐ Encourage student-led projects and real-world problem solving |
| 6. Multi-Stakeholder Collaboration | ☐ Involve youth, educators, developers, and policymakers in co-design of policies<br>☐ Promote shared governance principles and participatory frameworks<br>☐ Develop institutional policies with input from school boards, tech providers, and families |

### B. *Step-by-Step Guides*

#### 1) *For Educators and School Administrators*

Educators and school administrators play a critical role in shaping how young digital citizens interact with AI technologies. They should begin by thoroughly assessing AI-powered tools to ensure compliance with privacy laws and best practices. If required, parental or guardian consent should be obtained before introducing AI tools in the classroom. Privacy lessons should be integrated into the curriculum, teaching students about responsible data sharing and AI ethics. Additionally, educators should monitor how students interact with AI software, gathering feedback on usability and privacy concerns to inform future improvements.

#### 2) *For Parents and Guardians*

Parents and guardians must stay informed about the AI applications their children use, taking the time to review privacy policies and data-sharing terms. Setting boundaries around data-sharing practices is essential, encouraging children to question the necessity of certain permissions. Parental controls, whether built into the applications or through external monitoring tools, can provide an additional layer of oversight while respecting the child's autonomy. Ongoing communication about the reasons behind privacy settings is crucial, fostering a critical awareness of potential risks and empowering children to make informed decisions.

#### 3) *For AI Developers and Researchers*

AI developers and researchers have a responsibility to embed privacy considerations into the design and development of AI systems from the outset. This includes conducting privacy impact assessments to evaluate the ethical implications of AI on young users before deployment. Engaging with youth feedback during the design process ensures that AI features are relevant and user-friendly. Compliance with standards and regulations, such as OPC guidelines, PIPEDA, and GDPR, is essential to align development practices with legal and ethical requirements.

#### 4) *For Young Digital Citizens*

Young digital citizens should be encouraged to take an active role in managing their privacy. If unsure why an AI tool requests personal data, they should seek clarification from a teacher, parent, or trusted adult. Regularly reviewing and customizing privacy settings, such as camera or microphone permissions, helps maintain control over their data. Suspicious requests for excessive personal details should be reported to a trusted adult. By staying curious and critical, young users can develop a deeper understanding of how AI works and the implications of data sharing.

### C. *Proposed Policy Recommendations*

To further strengthen AI privacy governance, several policy recommendations are proposed. AI platforms targeting minors should be required to provide mandatory transparency statements, disclosing data flow, storage durations, and processing methods in plain language. Consent management systems should be implemented, offering tiered consent forms that balance parental oversight with youth autonomy. Privacy education initiatives should be integrated into school curricula, promoting awareness campaigns for both parents and students to ensure that young digital citizens are equipped with the knowledge to navigate AI-driven environments responsibly.

To support evidence-informed policymaking and enhance the accessibility of key recommendations, Table III summarizes the core governance levers explored throughout this paper. It presents a comparative view of their advantages, potential downsides, and implementation challenges. This format is intended to assist practitioners, educators, and policymakers in balancing design priorities, legal compliance, and stakeholder participation as they navigate the trade-offs inherent in building youth-centered AI privacy systems.

TABLE III
SUMMARY OF GOVERNANCE LEVERS FOR YOUTH-CENTERED AI PRIVACY

| Governance Lever | Key Advantages | Downsides / Risks | Implementation Challenges |
|---|---|---|---|
| Legal Standards (e.g., GDPR, PIPEDA) | Establish enforceable data protections; promote accountability | Often reactive; may lag behind technological innovation; may overlook youth-specific provisions | Varying jurisdictional scopes; slow legislative updates; inconsistent enforcement |
| Ethical-by-Design AI | Prevents harm at the design stage; improves system trustworthiness | May increase development complexity; lacks standardization | Requires technical expertise; industry-wide adoption remains uneven; complex to audit and certify |
| Privacy Education | Empowers youth through literacy and informed agency | Unequal access; curriculum integration can be slow | Requires educator training; curriculum development; policy support at multiple levels |
| Stakeholder Collaboration | Fosters shared accountability and inclusive governance | Risk of diffusion of responsibility; potential conflicts | Sustained engagement is resource-intensive; trust must be actively built |
| Youth Participation | Aligns policies with lived experiences and rights-based norms | Risk of tokenism or lack of real influence | Requires scaffolding, institutional mechanisms, and cultural shift |

### *D. Limitations and Future Considerations*

While these guidelines provide a robust framework for ethical AI governance, several challenges must be acknowledged. The feasibility of implementation varies widely across schools and AI startups, with differences in resources, technical capacity, and organizational maturity affecting the adoption of these recommendations. Smaller educational institutions or startups might lack the financial resources, expertise, or infrastructure necessary to fully implement robust privacy-by-design and cybersecurity measures. Additionally, achieving standardized compliance across diverse technological platforms and jurisdictional boundaries remains a substantial hurdle, complicating the effective enforcement of these guidelines. Table III also highlights the multidimensional nature of these trade-offs, illustrating how legal, technical, and participatory mechanisms, while conceptually sound, pose unique operational barriers depending on context and stakeholder capacity. This comparative view reinforces the need for adaptable, context-aware strategies that balance ambition with feasibility. Moreover, as AI regulations continue to evolve in response to technological advances and emerging ethical considerations, these guidelines will require periodic review and updates to remain aligned with new legislative requirements and societal expectations. A dynamic regulatory environment demands proactive adaptability, necessitating ongoing communication and collaboration among policymakers, educators, AI developers, and youth representatives. Maintaining a structured feedback loop with key stakeholders will be critical for capturing evolving concerns, technological shifts, and emerging best practices, thus ensuring that guidelines stay relevant and responsive to the rapidly changing AI landscape.

## V. CONCLUSION

The rapid integration of AI into digital ecosystems presents both opportunities and serious risks, particularly for young digital citizens who often lack the agency and skills to navigate complex data environments. This paper underscores the urgent need for proactive, stakeholder-driven privacy governance to ensure that AI systems adhere to ethical, transparent, and user-centered standards. Key areas requiring immediate attention include algorithmic transparency, privacy education, data ethics, and youth-specific cybersecurity protections. Addressing these challenges demands a multifaceted approach that blends regulatory enforcement, educational initiatives, and ethical AI design practices to foster a more accountable digital environment. Industry stakeholders must prioritize the implementation of clear, accessible explanations of algorithmic decision-making processes. Young users and their guardians should be able to understand how their data is being used. Privacy-by-design must become standard practice, minimizing data collection, applying strong encryption and anonymization techniques, and using age-appropriate consent mechanisms that are explicit and understandable. Equally critical is the rapid expansion of AI privacy literacy programs. Industry and education sectors must collaborate to embed digital ethics and privacy awareness into school curricula and public awareness campaigns, offering accessible tools for youth, parents, and educators. On the technical front, companies should adopt advanced, youth-focused cybersecurity strategies, such as differential privacy, federated learning, and blockchain-secured data exchange, to defend against evolving threats. These must be supported by mandatory breach reporting systems and enforceable accountability mechanisms. Furthermore, the legal landscape must evolve to hold companies accountable for privacy violations or unethical AI practices that affect young users. Policymakers must define clear responsibilities across the AI ecosystem and establish rigorous regulatory oversight, including meaningful penalties for non-compliance. Independent audits and third-party assessments can help build public trust by enhancing transparency and ensuring sustained accountability. Ultimately, empowering young digital citizens requires a coordinated effort, through structured education, enforceable privacy standards, transparent systems, and inclusive policymaking. The dynamic nature of AI demands sustained, cross-sector collaboration to ensure that technological progress aligns with ethical standards and social values, helping build a secure, equitable, and trustworthy digital future for all youth.